# Deep Neural Networks for Detecting Statistical Model Misspecifications. The Case of Measurement Invariance

Artur Pokropek[1], Ernest Pokropek[2]

[1]Institute of Philosophy and Sociology, Polish Academy of Sciences, Warsaw, Poland
[2]Division of Robotics, Perception, and Learning, KTH Royal Institute of Technology, Stockholm, Sweden

## Abstract

While in recent years a number of new statistical approaches have been proposed to model group differences with a different assumption on the nature of the measurement invariance of the instruments, the tools for detecting local misspecifications of these models have not been fully developed yet. In this study, we present a novel approach using a Deep Neural Network (DNN). We compared the proposed model with the most popular traditional methods: Modification Indices (MI) and Expected Parameter Change (EPC) indicators from the Confirmatory Factor Analysis (CFA) modeling, logistic DIF detection, and sequential procedure introduced with the CFA alignment approach. Simulation studies show that the proposed method outperformed traditional methods in almost all scenarios, or it was at least as accurate as the best one. We also provide an empirical example utilizing European Social Survey data including items known to be miss-translated, which are correctly identified with presented DNN approach.

Since the beginning of the modern social sciences, comparisons between countries, cultures, and communities were their constitutive part. Large portions of such comparisons focus on personality traits, attitudes, values, worldviews, and norms that are usually measured by multiple indicators (Davidov, Schmidt, Billiet, Meuleman, 2018). Such constructs are usually considered continuous latent variables, and the observed responses to questions are treated as manifestations of those latent traits (Brown, 2015; Joreskog, 1973). For example, the level of religiosity is usually measured by several questions on general religiousness and religious practices (e.g., Lemos, Gore, Puga-Gonzalez, & Shults, 2019). The attitudes to migrants are sometimes inferred by several items describing willingness to allow immigrants into the country (e.g., Davidov, Meuleman, Billiet & Schmidt, 2008), while the level of political participation is routinely identified by asking whether respondents had taken part in various forms of political activities (e.g., Kostelka 2014).

To assure the observed differences between groups reflect the actual differences between latent constructs and are not biased by artifacts induced by differences in measurement tools, the assumption of measurement invariance (or measurement equivalence) needs to be fulfilled - a property of a measurement, stating that the instrument measures the same constructs in the same way across various subgroups of respondents (Davidov et al. 2008). Mellenbergh (1989) pointed out that measurement invariance could be expressed as the assumption of conditional independence:

$$f(U|\theta, G = g) = f(U|\theta) \quad (1)$$

Measurement invariance holds when the probability of given response $U$ using response function $f$ given the latent trait $\theta$ is conditionally independent of group membership $G$ for given group $g$. In the context of cross-country comparisons, violation of this assumption is usually called measurement non-invariance, while in psychometrics, the term Differential Item Functioning (DIF) is more common (Holland & Wainer, 1993; Millsap, 2011).

Different nomenclature goes along with different problems and different methods of detecting violations of conditional independence. DIF methods were designed to cope with cognitive testing and psychological assessments where the number of groups is limited to two in the majority of the cases (usually gender or minority status), the number of items is significant, and violations of conditional independence are relatively rare, indicating some severe flaws in particular items (see Holland & Wainer 2012 for details). The research on DIF is rich and resulted in a series of well-established and well-evaluated methods (Angoff 1972, Lord 1980, Swaminathan & Rogers, 1990, Swaminathan & Rogers, 1990, Thissen, Steinberg, & Wainer 1993, Penfield, & Lam, 2000, Kopf, Zeileis, & Strobl 2015).

In cross-countries studies, settings are very different since the number of groups is usually significantly larger than two. For example, the European Social Survey (ESS) surveys samples from around 20 to 30 countries, depending on round (ESS 2020), while the newest seventh wave of the World Values Survey has conducted representative national surveys in as many as 80 countries (Hearpfer et. al. 2020). In 1984, the International Social Survey (ISSP) started with six countries and had grown to include 57 nations till 2020 (ISSP, 2021). The record in the number of participating countries belongs to the Gallup Global Well-Being Survey that in 2018 collected data from 144 countries (Wellcome Global Monitor, 2018), and these are just a few examples.

As extensive comparative surveys aim to ask questions concerning multiple topics, usually, the length of instruments used to measure individual constructs needs to be short to maintain acceptable interview time. For example, in round 9 of ESS (European Social Survey 2018), the average number of items per scale was just 4.6, while in the Survey of Adult Skills (OECD 2010), it was equal to 3.8. Moreover, in the context of cross-cultural studies, violation of measurement invariance (or DIF) is more of a rule than an exception (Davidov, Meuleman, Billiet & Schmidt 2008, Pokropek and Borgonovi 2017). In cross-cultural studies, respondents come from different countries, usually speak different languages, have been socialized in diverse cultural backgrounds, and might understand specific ideas or concepts in different, culturally-varying manners. Furthermore, the concepts measured like trust, values, attitudes, or political opinions are much more ambiguous and context-dependent than cognitive skills or personality traits (Davidov, Schmidt, Billiet, Meuleman, 2018).

In the context of cross-cultural studies, various methods were developed to test whether different assumptions of measurement invariance holds for the particular set of items (Kim, Cao, Wang, & Nguyen 2017). Those methods, sometimes called "classical measurement invariance" testing, could answer whether measurement invariance holds or not for a given set of items, but usually, we cannot point out specific non-invariant item parameters. Auxiliary methods designed to detect item non-invariant parameters were shown not to be very efficient (Finch, 2016; Kim et al., 1995; Magis et al., 2011; Penfield, 2001) and rarely tested (Davidov et al. 2014, Fitzgerald & Jowell 2010, Skjak 2010, van de Vijver 2011). This is a frustrating gap, especially in the light of the fact that recent latent variable models can maintain comparability of results but only if non-invariant parameters are correctly identified (Byrne, Shavelson & Muthén 1989; Steinmetz 2013, van de Schoot, Kluytmans, Tummers, Lugtig, Hox, & Muthén 2013, Asparouhov & Muthén 2014; De Jong et al. 2007, Muthén & Asparouhov, 2012, 2014). In other words, we have tools to fix some of the problems of comparability, but we do not know where to apply them.

To solve this problem, we are applying a methodology based on Deep Neural Networks (DNNs). This approach has shown notable success in anomaly detection problems in industrial (e.g., Staar, Lütjen & Freitag 2019), medical (e.g., Anwar, Majid, Qayyum, Awais, Alnowami, & Khan, 2018), or cyber security (Ferrag, Maglaras, Moschoyiannis & Janicke, 2020) settings. We propose to look at the problem of measurement non-invariance of items or DIF as an anomaly detection task that could be handled by DNN combined with a simulation approach. The idea is to train a DNN in a controlled environment provided by simulation data and teach it to detect non-invariance instances in real scenarios. In fact, from a statistical point of view, there are no contraindications to treat non-invariant items in cross-country studies similar to broken screws in mechanical devices, brain tumors in magnetic resonance imaging scans, or wire transfer frauds in finances.

The contribution of this paper is threefold. First, we review and discuss existing methods for detecting non-invariance item parameters (or DIFs) that could be used in settings of cross-country group comparisons. Second, we introduce a new method for detecting non-invariance in item parameters, test it in a simulation study and present an application on a real data example. Third, we confront newly proposed approaches with the existing ones, contributing to the ongoing research by validating different methods under various new conditions.

## Existing methods for detecting non-invariant items in multiple-group settings

There are several methods for detecting item non-invariance; however, not all of them are suitable for multiple-groups settings. For example, the most popular methods based on generalized logistic regression approach (Magis et al., 2011) are designed to detect non-invariant items, but without specifying in which groups the invariance holds, and in which it does not, just flagging the particular items as non-invariant between all of the groups. Some of the methods like Mantel–Haenszel procedures (Penfield, 2001) cannot detect non-invariance in item loadings (also called non-uniform Differential Item Functioning; DIF). Most of the methods, even those with the word "multiple-groups" in their name, were designed and tested for 3 or 4 groups (Fidalgo & Scanlon 2010, Woods, Cai, & Wang 2013). Other methods like random item effects models (De Boeck 2008; Fox, 2010) require a huge number of groups (more than 40 according to some accounts; Meuleman & Billiet, 2009) that could be suitable only for the most extensive comparative programs. For this article, we have chosen methods governed by the three criteria: (1) method could be applied to moderate and large numbers of groups, (2) provides information of non-invariance for both item parameters (thresholds and loadings), or in other words, could detect both uniform and non-uniform DIF and (3) provides information on item non-invariance for each group. We have chosen two approaches from the latent variable framework: CFA/IRT model misfit indices, sequential item parameters comparison implemented in the alignment method, and one approach based on observed scores: pairwise logistic regression DIF detection.

### Methods of detecting non-invariance items in the latent variable framework

The most popular approach to address the problems of group comparisons and potential compatibility problems in cross-country settings is the latent variable framework (Davidov et al., 2008). In this paper, we will restrict our investigation to the case where indicators are binary, and therefore multiple group Item Response Theory (IRT) models will be used (Lord, 1980; W. J. van der Linden & Hambleton, 2013). However, the approaches we discuss here are readily applicable to continuous (Brown, 2015; Joreskog, 1973) and polytomous indicators (W. J. van der Linden & Hambleton, 2013). Some authors refer to IRT models as CFA models with categorical indicators. IRT and CFA models for categorical data are essentially the same models with some minor differences in parametrization. In this paper, we are treating IRT and CFA models for categorical data as the same models.

Let $U_{ipg}$ be a dichotomous indicator (response to the item) for a latent factor $\theta_{pg}$. According to the IRT modeling, the probability of responding positively on item $i$ ($U_{ipg} = 1$) is given by the logistic model:

$$P_{ipg}(U_{ipg}|\theta_{pg}) = P_{ipg}(\theta_{pg}, G = g) = \frac{1}{1+exp\,(\tau_{ig}-\lambda_{ig}\theta_{pg})}\ , \theta_{pg} \sim N(\alpha_g, \psi_g^2) \quad (1)$$

where $i$ = 1,…,$I$ denotes the item index, $p$ the person index and $g$ = 1,…,$NG$ the group index, , $\tau_{ig}$ and $\lambda_{ig}$ are the item parameters, threshold (in some contexts named intercepts), and loading respectively, $\theta_{pg}$ is a factor reflecting the latent trait that is assumed to be normally distributed in each group. The graphical representation for four groups is presented in Figure 1.

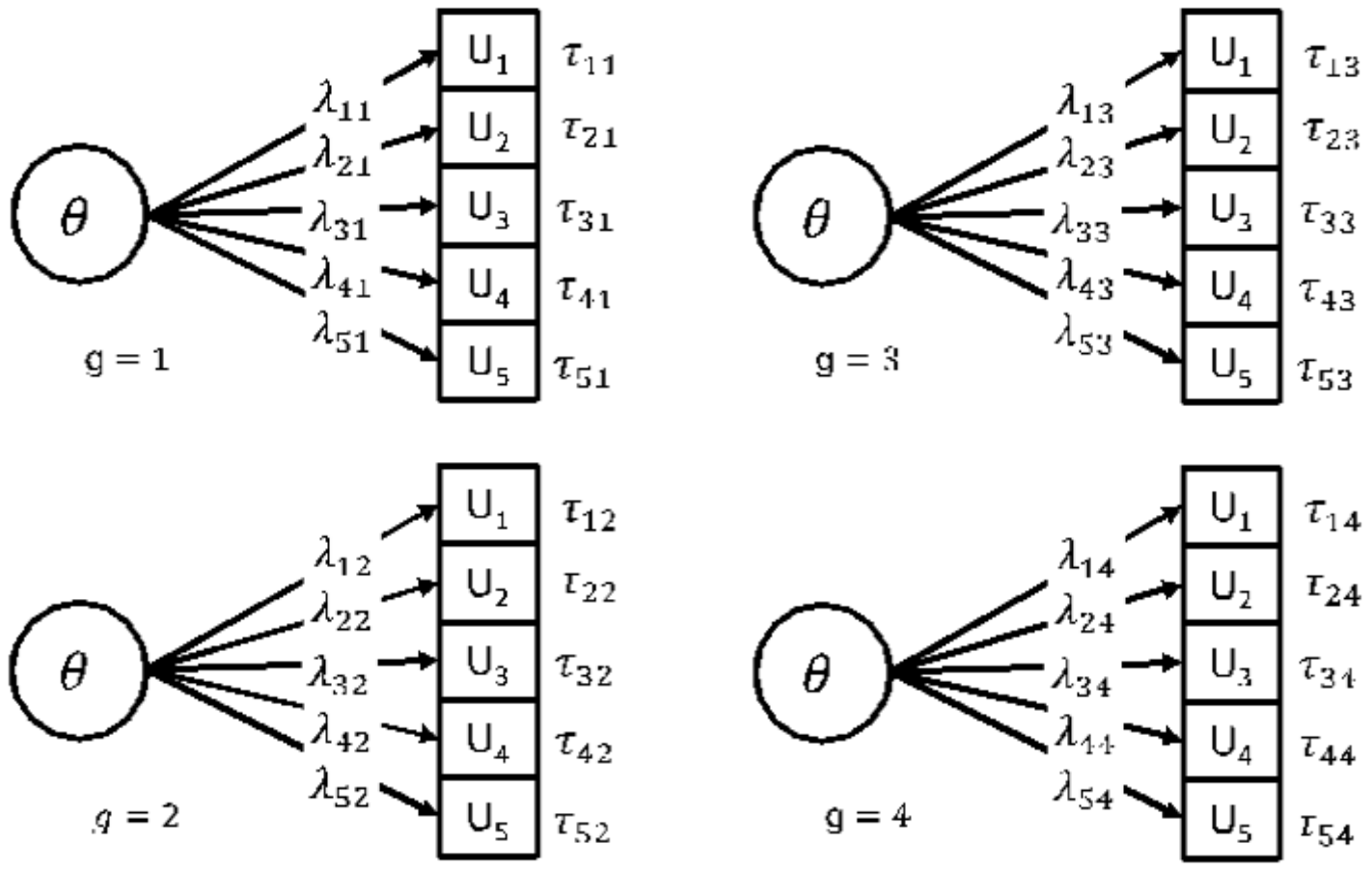

*Figure 1.* Graphical representation of four group IRT model

Unless some restrictions on parameters are imposed, the multiple group IRT model is not identified because not all item loadings and item thresholds can be simultaneously estimated along with group means $\alpha = (\alpha_g)$ and standard deviations. The standard approach in multiple-group settings is to constrain all item parameters to be the same in all groups and fix the mean together with the variance of one of the groups to 0 and 1, respectively. Alternative identification strategies are possible, but this is the most common in multiple-group settings (Svetina, Rutkowski & Rutkowski, 2020). This model is often referred to as *a scalar* model and assumes full-invariant item parameters (loadings and thresholds).

Assuming conditional independence of the responses, the models could be estimated using various techniques, including maximum marginal likelihood (Bock & Zimowski, 1997), weighted least squares estimations and its extensions (Beauducel & Herzberg, 2006; Li, 2016) as well as Bayesian estimation (Fox, 2010).

**Item fit statistics of scalar model.** The popular approach for detecting item non-invariance is based on assessing the so-called modification indices criteria (MI, also referred to as the univariate *Lagrange multiplier*; Sarris, Satorra & Sorbom 1987, Sorbom, 1989). The modification index expresses an approximation of how much the overall model $\chi^2$ would change when the parameter is freed from a constraint. In the context of a multigroup model this means that item parameter is estimated separately instead of being constrained to be equal in all groups. Following statistic, the expected parameter change (EPC) index is strictly related to MI. EPC signals the value of the expected parameter change between the fixed and freely estimated parameter. In the context of measurement invariance, EPC expresses the difference between fixed item parameter estimation across groups and free estimation of this parameter for a particular group (Kaplan 1989; Whittaker, 2012). The EPC is, in fact, directly analogous to MI but reflects the approximate changes in metric of the parameters, which might help experienced scientists to interpret it better and formulate more meaningful cut-off criteria. Saris et al. (2009) noticed, the EPCs are consistent estimates of the actual value of the parameter, and MI gives a good indication of local misfit, provided that the other restrictions in the model are (approximately) correct, which was confirmed in simulation studies (MacCallum 1986, MacCallum & Austin 2000).

Described fit statistics have some problems. First, no established thresholds for those measures exist, and researchers rely on subjective decisions in operational use. Second, the consistency of the indices, that is, their increasing ability to produce correct inferences with the increased sample size, depends on the model's overall fit to the data (Saris et al. 2009). Third, those fit statistics work well for extensive non-invariance violations, but when differences between latent means of the groups are not significant, the power of DIF detection decreases significantly. This was previously noticed by Oort (1998) in a simulation study performed using two group scenarios investigating MI and EPC and was later confirmed in multiple-group settings and for other similar approaches (Buchholz & Hartig 2020). Moreover, it was shown that in situations where latent means are located far from item parameters, the sensitivity of detecting non-invariance of items is very low (Tijmstra, Bolsinova, Liaw, Rutkowski & Rutkowski 2020).

**Alignment sequential DIF algorithm.** The alignment approach was proposed by Asparouhov and Muthén (2014) and then further developed by other researchers (Robitzsch 2020; Pokropek, Lüdtke, and Robitzsch 2020). It aligns item parameters from group-specific configural CFA or IRT models (where item parameters are estimated without equality constraints across groups) into the most optimal invariance pattern that allows the estimation of group-specific factor means and variances without requiring the exact measurement invariance. It is doing that by determining α and ψ in such a way that the amount of measurement non-invariance is minimized. This is done by an alignment function that optimally aligns group-specific item parameters, and such procedure consists of two steps.

First, configural measurement models are estimated for each group, where each model is identified by setting the mean to zero and the standard deviations to one while all item parameters are estimated freely in each group. Because the means and standard deviations of latent variables are functions of item parameters:

$$\lambda_{ig} = \frac{\lambda_{ig,0}}{\psi_g} \quad \text{and} \quad \tau_{ig} = \tau_{ig,0} - \frac{\lambda_{ig,0}}{\psi_g}\alpha_g \ , \tag{2}$$

where $\lambda_{ig,0}$ and $\tau_{ig,0}$ are loadings and thresholds from a configural model, respectively, shifting values of means and standard deviations of the group would result in changes of item parameters. This relation is used in the second step of the alignment algorithm, where the procedure searches for the configuration of means and standard deviations that would minimize the difference between item parameters between groups. To achieve this, Muthén and Asparouhov (2014) proposed to use a loss function that would be the same for loadings and thresholds:

$$f = \sqrt[4]{x^2 + \epsilon} \tag{3}$$

where $x$ is the difference between item parameters and $\epsilon$ is a small number such as 0.0001 to make the function differentiable. The alignment procedure penalizes differences in item thresholds and item loadings between groups and hence minimizes the extent of measurement non-invariance according to the loss function. There are various possibilities of specifying the implementation of the alignment optimization. In our application, we used a fixed version with the default settings of Mplus Muthén and Asparouhov (2014).

For detecting non-invariant item parameters, an iterative pairwise procedure is used after the alignment estimation is completed. This procedure is conducted for each item parameter separately. In the first step, for each item parameter, the largest invariant set of groups is determined. The group is found by conducting pairwise tests for each pair of groups on each parameter. If the differences between item parameters are not significant ($p>0.001$), pairs of parameters are connected. The largest connected set of groups is considered as the starting invariant set. In the second step, the average of each parameter from the starting invariant set is computed. Then the parameter from each group is compared to the average of the parameters in the invariant set. If there is a significant difference, the group will be removed from the invariance set. Otherwise, this group will be added to the invariance set. The second step is repeated until a stabilized invariance set has been found. Groups that were not assigned to the invariance set are considered to be non-invariant in that parameter (Muthén and Asparouhov 2014).

Alignment is one of the best-studied procedures of detecting non-invariant parameters in the context of studies with large numbers of groups: Asparouhov and Muthén (2014) conducted a series of simulation studies to evaluate the quality of the alignment method analyzing scenarios with five items with a number of groups set at 2, 3, 15, and 60, and group size set at 100 or 1,000. Finch (2016) conducted a simulation study to compare the alignment method with the Generalized Mantel-Haenszel method, generalized logistic regression, and the Lord's chi-square test on detecting uniform DIF across multiple groups. Twenty dichotomous items with 2, 3, and 6 groups were employed. Lin (2020) conducted a study with 20 items in 24, 40, and 80 groups with sample sizes of 100 and 500. The picture is consistent: simulation studies showed that the detection rate was very high for thresholds, at least as good as for well-established methods for two groups. The detection of non-invariance of loadings was substantially worse than thresholds but acceptable in situations where the means of the groups were equal. Interestingly, in conditions with unequal means, alignment methods would lead to incorrect DIF detection (Lin 2020). This, not a widely commented fact, is especially problematic for cross-country comparisons where cross-country differences are very plausible.

**Logistic regression approach.** In practice, there are no established algorithms to detect non-invariance in multigroup settings for observed scores methods. However, as most of them showed good performance in comparing two groups, for multiple group cases one could employ this pairwise methodology in a sequential manner. As we have not found any established algorithms, we propose an *ad hoc* procedure for this paper. We decided to use logistic regression for detecting non-invariance as proposed by Swaminathan and Rogers (1990) because of the good performance of this method in two group settings and straightforward implementation.

In the proposed approach, responses to the items are set as dependent variables. To the baseline model sum of item responses reflecting measured construct are added as independent variables:

$$ln\left(\frac{\pi_{pg}}{1-\pi_{pg}}\right) = \alpha + \beta_1 S_{pg} \tag{4}$$

where

$\pi_{pg}$ is a probability of a correct response for person $p$ in group $g$,

$S_{pg}$ is a score, sum of item responses as a proxy for latent variable,

$\beta_i$ identifies the relation between correct response and observed sum score,

In the model used to detect uniform DIF binary group indicator is added as second independent variable:

$$ln\left(\frac{\pi_{pg}}{1-\pi_{pg}}\right) = \alpha + \beta_1 S_{pg} + \beta_2 G_g \tag{5}$$

where

$G_g$ is a binary indicator of a group membership,

$\beta_2$ identifies uniform DIF effect,

For detecting uniform DIF log-likelihood ratio test is used comparing the baseline model with the model with the group indicator described in equation (5). For detecting non-uniform DIF the model is extended with interaction term:

$$ln\left(\frac{\pi_{pg}}{1-\pi_{pg}}\right) = \alpha + \beta_1 S_{pg} + \beta_2 G_g + \beta_3 (S_{pg} * G_g) \tag{6}$$

where

$(S_{pg} * G_g)$ is an interaction variable,

$\beta_3$ identifies non-uniform DIF effect.

A likelihood ratio test that compares model (5) with (6) identifies non-uniform DIF. This approach allows only for pairwise comparisons. Each group has to be compared with all others, and based on those multiple comparisons, conclusions about comparability must be drawn. Here, we are proposing a simple algorithm. In the first step, pairwise DIF analysis between all groups separately for all items is performed. The likelihood-ratio test with Benjamini-Hochberg adjustments for multiple comparisons (Benjamini & Hochberg, 1995) is performed for each comparison. The p-value of the log-likelihood ratio test is computed. For each item in each group, the number of DIF effects in pairwise comparisons against other groups is counted, where the DIF is considered significant for $p < 0.01$. When the proportion of DIF effects in comparisons for an item in particular groups is more prominent than half of the groups, we flag that item as non-invariant. Uniform and non-uniform DIF effects are computed in different comparisons. As mentioned above, this procedure was invented *ad hoc* because of a lack of established alternatives. However, the initial tests and final results showed that the procedure is very effective and can compete with more sophisticated approaches described earlier.

It is worth noting that all of the methods presented in this section and the new method proposed in this paper assume the most fundamental level of measurement invariance. This configural invariance assumes the same dimensional structure but different item parameters. The configural invariance is being tested using fit measures and acceptable fit threshold for each of the groups (for details, see Byrne et al. 1989, Meredith 1993). In this paper we investigate the performance of the models assuming that this basic assumption is fulfilled, and for real-life examples, we have chosen the data where configural invariance was proven to hold (Koc 2021).

## Proposed Approach

The proposed approach is similar to the idea proposed by Zhang and colleagues (2017). This procedure consists of four steps. The first one involves examining the structure and features of the target data, that is, the dataset where measurement invariance needs to be tested. In the second step, Monte Carlo simulations generate data reflecting measurement situations and potential ranges of non-invariance. Thirdly, the generated data, sets of responses for all items and all groups, are used to train DNN. Finally, in the fourth step, trained DNNs are used to predict item non-invariance in the target dataset.

### Step 1 Initialization

The proposed approach starts with an investigation of the target dataset. This is done to establish the dimensions of data structures (number of items, number of respondents, and groups) and primary boundary conditions for Monte Carlo data generation in Step 2. For the measurement invariance, the crucial element is the distribution of item and group parameters. To approximate actual item parameters distributions, configural models could be run (items parameters estimated free in each group). To obtain the distributions for latent means and standard deviations scalar model (assuming non-invariance) could be estimated. Of course, the estimated parameters would be biased in the presence of non-invariance. However, they would give reasonable boundaries for Monte Carlo data generation where simulated item parameters would be drawn for each iteration. In this step, the size or sizes of misspecifications to be detected need to be determined. The good tool should detect non-invariance in parameters that are substantially important and not driven by random processes. These values could be based on previous simulation studies (e.g., Pokropek, Davidov & Schmidt 2019; Pokropek, Schmidt & Davidov 2020), sensitivity analysis (Kuha & Moustaki 2015), or other methods like effect size indices for measurement invariance (Nye & Drasgow, 2011, Gomer, Jiang & Yuan 2019).

### Step 2 Monte Carlo data generation

The Monte Carlo data generation step consists of three phases. In the first one, a multigroup measurement model is defined, and distributions of parameters are provided together with distributions for size and allocation of non-invariance based on the outcomes of step 1. For the next phase, item parameters (item thresholds and loadings) and non-invariance values (local misspecifications for thresholds and loadings for each group) are sampled from distributions defined in the first phase. In the third and final phase, one dataset is generated containing item responses and group membership (based on sampled parameters). Together with the data matrix, a vector containing binary indicators for item bias is produced. The vector is constructed so that it refers to each parameter in each group and takes a value of 0 for invariant parameters and 1 for non-invariant parameters. Once the data matrix and bias indicator vector are created and stored, we head back to phase two and continue until sufficient data sets are generated. Empirical checks showed that for the applications presented in this paper 300 000 datasets and invariance vector indicators are sufficient for good detection of non-invariance parameters.

**Step 3 Deep Neural Networks Learning**

In the third step, generated datasets are used to train DNN. An artificial neural network is a computing system consisting of (artificial) neurons formed in layers and connected. A simple way to understand the concept behind this model is to think about a single neuron (described in detail in the next paragraph) performing a similar task as a logistic regression unit, i.e., computing a weighted average of its inputs and passing the result through a function limiting its output. Intuitively, if we increase the number of such decision boundaries by utilizing many neurons and connecting them, we might be able to tackle more complex, non-linear problems; as the universal approximation theorem states, such a network with only a single layer can approximate any function (Goodfellow et al., 2016). The previously mentioned weights, initially set randomly, change their values in order to minimize the error of prediction of the whole network during a process called training; thus, the final model can be simply treated as a function which computes a set of weighted averages of the inputs consecutive neurons to produce the prediction.

The first layer is called the input layer, as it receives the features. Intuitively, the last one is called the output layer, which corresponds to the model's prediction. The layers between these two are referred to as the hidden layers, and when the neural network consists of more than one of them, it is classified as a deep neural network (DNN). The neurons from one layer are connected to other neurons from adjacent layers, and each connection has a weight assigned to it, representing its relative importance. The number of such weighted connections may vary depending on the architecture of the DNN. Each neuron has its internal state, called activation - when the *n*-th neuron in the layer is active, it generates an output $z_n$, which then traverses by its output connections to the next layer and might be defined as:

$$z_n = h(\sum_{i=1}^{N} (w_i * x_i)) + b) \quad (7)$$

Where $h$ denotes the activation function, $N$ the number of neurons in the previous layer (or the number of inputs to the model), $x$ and $w$ being $i$-th input and weight consecutively, and $b$ being the optional bias (in our study, b=0 for all models in all layers). Then, the vector $[z_1, z_2, \ldots, z_n]$ becomes the input for the next layer or represents the predictions (if the layer is the output one). The graphical representation of equation (7) is depicted in Figure 2.

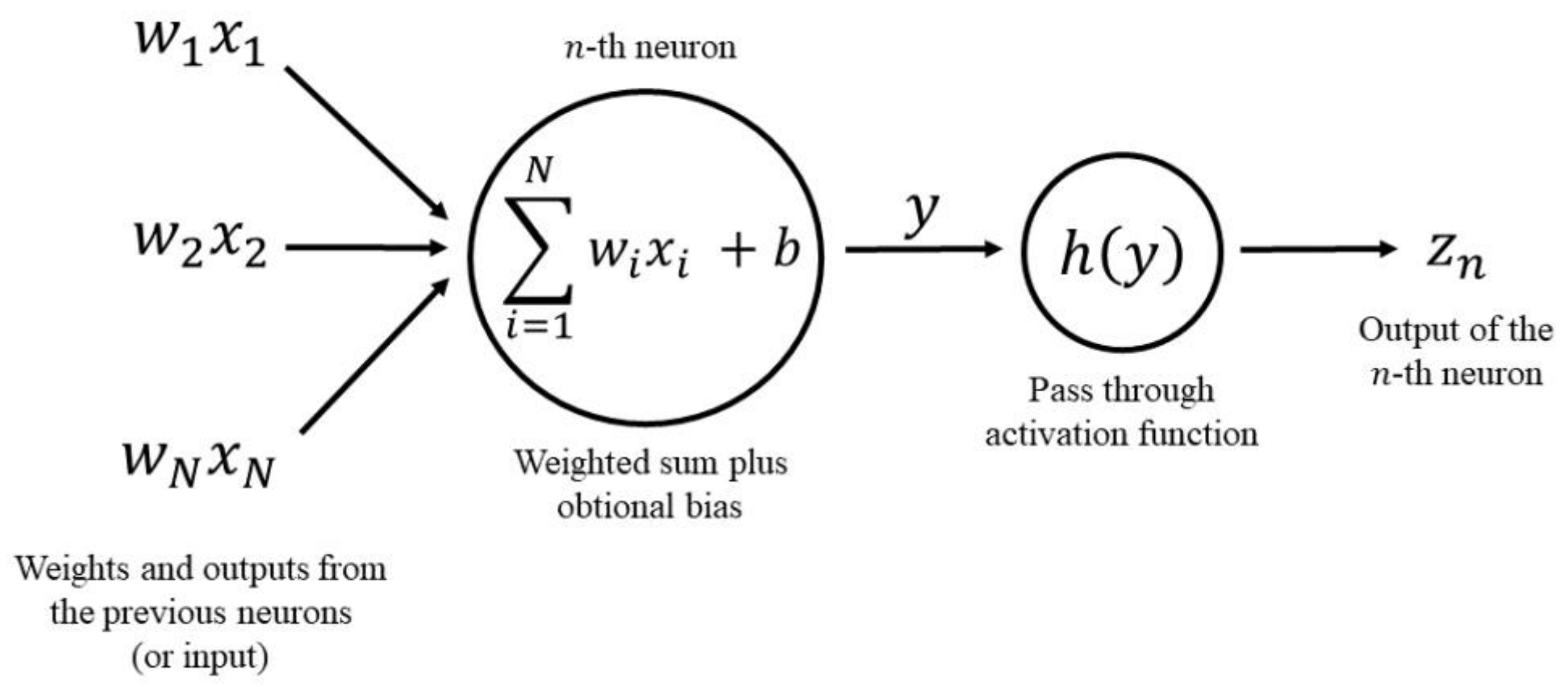

*Figure 2.* Graphical representation of the neuron in DNN and its relation input and output elements of the neural network

Currently, one of the most popular activation functions is the rectified linear unit (*ReLU(z) = max(0, z)*) due to its processing speed and relative resistance to large input values. Contrary to, for example, logistic function or hyperbolic tangent, ReLU has no maximal value, which helps to reduce issues during DNN estimation (Geron, 2019).

Calculating weights of connections between neurons, which are commonly initialized randomly, is called training. As in all machine learning algorithms, this task consists of numerical optimization, i.e., a minimization of the loss function often referred to as cost function, or in statistics, a likelihood function. The loss function is a metric that defines to what extent the prediction differs from expected values; for example, a simple yet powerful statistical metric such as the mean squared error is often used for this purpose. The loss function is also used in the weight optimization technique called backpropagation (Rumelhart et al., 1986) - the error at the ANN's output is measured relative to the influence of the neurons of the previous layer, and this step is repeated until reaching the input layer. This influence is calculated using a gradient of the loss function, associating it with a given state of the neuron, and finally, the weights are updated to minimize the loss function. The most popular optimizers used for modern-day NNs are stochastic gradient descent (SGD) with its improvements (momentum, Polyak, 1964; Nesterov, 1983), as well as adaptive moment estimation (Adam, Kingma & Ba, 2017), or adaptive subgradient (AdaGrad, Duchi et al., n.d.). Prediction in DNN may take different forms depending on the problem. For example, having a binary classifier (predicting a 0 or 1 outcome), we could rely on activation of the output neuron (active - non-active) or on the value of its output, i.e., magnitude of its activation function.

Elements that affect the training process, such as the number of layers, number of neurons per layer, and parameters of the optimizer, are called hyperparameters, and optimizing them to match the best performance of the NN is called tuning. For the whole process of model's training and evaluation, the data has to be split into:

(1) the training set, used for learning purposes of the model,

(2) the test set, which is the data that has not been present to the model during the training process to later evaluate it on, and eventually the

(3) validation set; popular mostly in DNNs (used to monitor the model's performance during the training process).

This division, often referred to as train-test split or train-test-validation split, usually implies the majority of the samples to be put into the training set (about 60 to 80% of the whole data). Although one may decide not to select any data for validation and focus on the metrics obtained from the trained model on the test set, having a validation set is helpful for techniques such as early stopping, when the training process is automatically stopped after meeting some criteria, such as a drastic increase in the loss function for validation data. Such an event may signalize that the model is too optimized to fit the training data.

This phenomenon is called overfitting, and it is a significant issue of training complex machine learning models, such as DNNs. It is a good practice to fight overfitting by manipulating the architecture of the model - for DNNs, the most widely used techniques to prevent such phenomenon are dropout (Hinton et al., 2012; Srivastava et al., 2014) and batch normalization (Ioffe & Szegedy, 2015). Dropout works by randomly selecting a fraction of weights on the layer that should be ignored during the training phase, making them not considered during any pass of the training data throughout the DNN (this does not have an effect when making predictions). This yields in some of the neurons being 'inactive' during a particular training phase, so they are effectively skipped. This procedure reduces the capacity of the DNN by introducing noise to the training, as with each pass of the data, the number of active connections between the neurons varies. This may be compared to creating many layers in the DNN model and then producing a final one by averaging the result.

As the data sets for deep learning are usually large, they are usually split into batches - parts that are used for learning, i.e., optimization of weights of DNN. The batch might be referred to as a fraction of the whole training data. Furthermore, a process of the batch passing through the DNN forward and backward is called an epoch. Batch normalization for a given layer subtracts the mean from its output and divides the result by its standard deviation. It has been proven that batch normalization vastly speeds up the training process (Ioffe & Szegedy, 2015).

Two DNN models were trained for the proposed application based on generated step 2 item responses and invariance indicator vector: one for non-invariant thresholds and the second for non-invariant loadings. Given that neural networks directly benefit from larger datasets (Alom et al., 2019), DNNs architecture that was tested for the proposed application consists of 8000 neurons in the first layer (exposed to various numbers of inputs, in our case - item responses), two hidden layers with 8000 neurons each, and the output layer. Due to the measurement non-invariance detection problem being somewhat difficult to solve using linearly separable functions alone, we decided not to rely on 'shallow' ANN models and utilize the deep learning approach (Reed & Marks, 1999). Given the complexity of this problem and a relatively large number of inputs, the model had to consist of many neurons in the hidden layers to provide a proper generalization (Goodfellow et al., 2016). Although there is no general rule nor an analytical method for providing the most optimal values of the hyperparameters as mentioned earlier (Goodfellow et al., 2016), we experimented with various promising architectures following the previously stated assumptions. The presented architecture has been found as the most effective one for our study given its performance, as we conducted an effectiveness analysis based on the validation set, tuning the hyperparameters when necessary (which is typical conduct for DNNs). For all layers except the output one, ReLU activation has been used, whereas the neurons responsible for prediction used the sigmoid function (producing the probability of activation, which was necessary for multilabel binary classification tasks we faced). Batch normalization was used for the first layer before dropouts not to introduce variance shift (X. Li et al., 2018). Both of the hidden layers are influenced by 50% dropout, reducing the complexity of the DNN during the training without oversimplifying the model. The graphical representation of used architectures can be seen in Figure 3. Please note that we used the same DNN architecture for all the presented studies, except the shape of the input and the number of neurons in the output layer. This is also our recommendation for further studies.

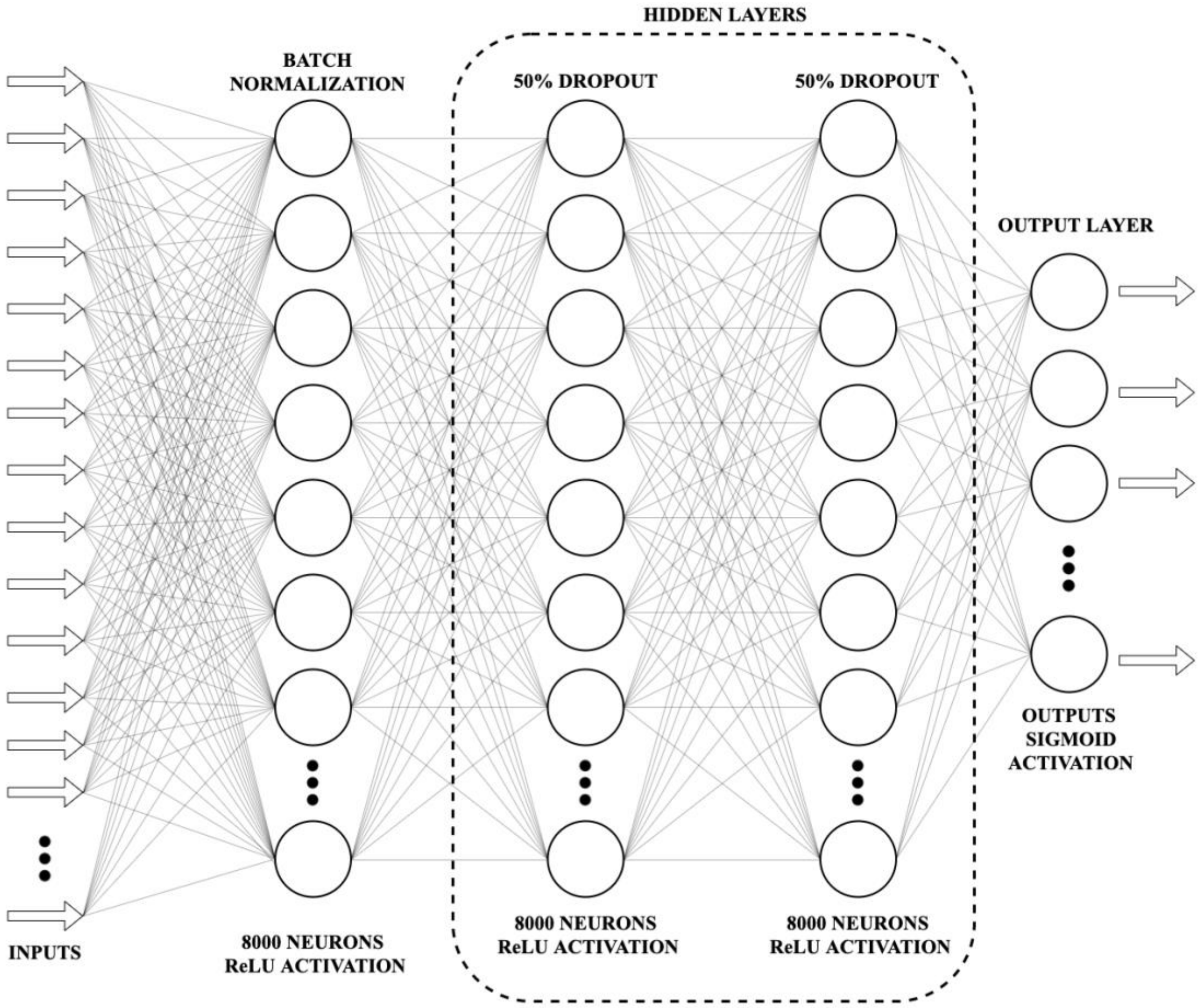

*Figure 2.* Graphical representation of the architecture of used DNNs

Nesterov momentum optimization (Nesterov, 1983) has been used to optimize the loss function, for our case binary cross-entropy (also referred to as *logloss* function), given with the following formula:

$$L(\hat{y}_i, y_i) = -\frac{1}{N}\sum_{i=1}^{N} \left[y_i log(\hat{y}_i) + (1 - y_i) * \log(1 - \hat{y}_i)\right] \quad (8)$$

Where N is the number of predictions, $y_i$ is the i-th actual value, and $\hat{y}_i$ is the i-th predicted value. The learning rate was equal to 0.1 and momentum to 0.8. This high learning rate was proven effective for our study due to the complexity of the input data and due to the use of batch normalization, yielding improved generalization (Lewkowycz et al., 2020) as expected. With batch size equal to 256, approximately only three epochs were sufficient to train the model predicting the thresholds, and 8 to obtain satisfying metrics for the loadings concerning all the studies.

**Step 4 Deep Neural Networks Predicting and Non-Invariance Flagging**

Once the parameters of DNN are estimated, the model can predict the outcomes from new data. Based on the trained models and input data for each item received, the probability of violation of conditional independence is computed. For each of the simulations, the input to the DNN is a one-dimensional vector containing all item responses from a sample (or all) of respondents. Passing through the model, the output layer represents a one-dimensional vector of probabilities (achieved by using sigmoid activation) of the presence of either non-invariant thresholds or non-invariant factor loadings.

In different situations, different probability thresholds could be used. We propose establishing cut-offs based on the investigation of ROC (Receiver Operating Characteristic) curves to determine the optimal balance between true-positive and false-positive rates of the detection in particular situations. If there are no solid objectives for preferring bigger rates of either true-positive or false-positive rates, optimal thresholds could be used. In the presented applications, we used a method proposed by Liu (2012) to estimate the optimal cutpoint that maximizes the product of sensitivity and specificity.

Suppose the research community would successfully adopt the approach, and various DNNs would be trained and made available. In that case, an applied researcher could only perform the last step using trained networks produced by the specialists in machine learning and statistics by choosing the one with the best-documented performance and mostly suited structure for selected problems. Generally, this would not be a big difference from the situation where applied researchers use statistical software written by well-trained specialists from statistics.

As authors understand that the readers might not be familiar with the utilization of deep learning in their studies and the presented description of this approach might seem intimidating, the list below introduces various software solutions with a relatively low entry threshold for setting up the training environment.

1) **Keras** module for **TensorFlow** (Python3[1] & R[2]); used in this study, simple and straightforward package for developing code for creating and training artificial neural networks

---

[1] https://keras.io/

[2] https://keras.rstudio.com/

2) **Deep Learning Toolbox**[3] for MATLAB; a complete framework for developing and evaluating deep architectures with many features such as Deep Network Designer, barely any previous coding experience is necessary
3) **Azure Machine Learning**[4] from Microsoft; a standalone online platform for developing and evaluating various machine learning models, including deep neural networks, with features such as Automated ML, which requires only supplying the data from the user

Additional reading on deep neural networks might be found in Deep Learning (Goodfellow et al., 2016) and Neural Networks and Deep Learning (Nielsen, Michael 2015.) books, available online, providing extensive explanations of the underlying mathematical concepts. For a more hands-on experience, we refer the reader to the official documentation of selected software solutions.

## Simulation Study

In this section, we would like to show how the proposed approach behaves compared to traditional methods. We designed this study to reflect the real-life analysis situation with many groups, a small number of items with substantial variation between latent means, and different sizes of parameters non-invariance. The conditions seem natural, but surprisingly, the literature review shows that developed methods for handling many groups do not perform optimally in such settings (Pokropek et. al. 2019). We expect that our method would outperform previous approaches in all conditions, but discrepancies will be the largest for conditions with the most significant number of groups and small numbers of items in scale: condition 4 with ten groups and three items per scale. In order to do it, a simulation study has been conducted with its architecture, as provided in Table 1.

Table 1 *Simulation architecture for four conditions*

| | Condition 1 | Condition 2 | Condition 3 | Condition 4 |
|---|---|---|---|---|
| Groups no. | 4 | 10 | 4 | 10 |
| Observations per group | 400 | 400 | 400 | 400 |
| Indicators no. | 5 | 5 | 3 | 3 |

We based our simulations on empirical ESS data. We performed the analysis advised in step 1 of the new approach and investigated the empirical distributions of item parameters from the political participation scale of the 7th round of ESS (ESS 2014). We ran the MG-CFA configural model using five binary items designed for measuring political participation and investigated the distribution of item parameters. The distribution of parameters from ESS could be approximated by a skewed normal distribution with a mean of 2.95, a standard deviation of 1.22, and a skewness parameter of 10 for loadings and characterized by a skewed normal distribution with a mean of -0.21, a standard deviation of 10.64 and a skewness parameter of 0.72 for thresholds. To fit better to empirical results, we additionally truncated the distributions of factor loadings to range between

[3] https://www.mathworks.com/help/deeplearning/ug/deep-learning-in-matlab.html
[4] https://azure.microsoft.com/en-us/services/machine-learning/

1 and 7 and thresholds to range between -2 and 1. A Scalar multiple-group CFA (MG-CFA) model was used for establishing distributions of latent means. The good approximation of obtained group means was a normal distribution with zero mean and standard deviation equal to 0.2, and a good approximation of within-group standard deviations was provided by a uniform distribution bounded by 0.75 and 1.25. Having established the possible distribution of parameters, we generated the data.

Unique sets of item parameters, group means, and standard deviations were sampled from the distributions described above in every single replication. Then data were generated for each group assuming measurement invariance, i.e., the same item parameters for each group but different means and standard deviations of the construct. Next, randomly sampled parameters were modified to reflect a violation of the invariance assumption. Each of these parameters was changed by adding or subtracting a value that was sampled from a uniform distribution bounded by 0.3 and 1.0 for factor thresholds and 0.3 to 2.0 for factor loadings. Each parameter was randomly assigned to have a positive or negative sign of a modification. The lower bound of the modification describes medium non-invariance that moderately changes the results, while the upper bound represents a considerable magnitude of non-invariance. Those values are consistent with previous literature and simulation studies on measurement invariance and DIF (Kim, Yoon & Lee 2012, Woods 2009, Stark, Chernyshenko, & Drasgow, 2006, Flake & McCoach 2018).

Different values of item parameters and item non-invariances were sampled from the defined distributions in each simulation dataset. In each replication, the number of non-invariant groups was sampled (with the restriction that at least one group should be non-invariant). Also, the number of non-invariant items in each group was randomly assigned for each replication with a minimum number of 1 and maximum of 3 (in 5 items conditions) and a maximum of 2 (in 3 items conditions). For each simulation, 5000 data sets were generated (the code for data generation is available in online supplementary material). On each dataset, five methods of invariance detecting were performed: (1) modification indices (MI), (2) expected parameter change (EPC) - both of the indexes were obtained from a CFA/IRT scalar model, (3) logistic regression approach (Logistic), (4) alignment sequential DIF algorithm (alignment), and (5) DNN approach (DNN). For the DNN we generated separate additional 300 000 data sets for training the model. To obtain estimates for methods (1), (2), and (4) Mplus version 8 computer program was used (Muthén & Muthén, 2017). For (3), we implemented a custom algorithm using the *difLogistic* function from the difR R package (Magis et al., 2010). For (5) we used Keras' sequential model, provided from the TensorFlow 2 package for Python3 (Abadi et al., 2015).

## Simulation Study Results

We present the results using ROC curves for MI, EPC indices, and predicted probabilities of DNN. As approaches based on logistic regression and alignment optimization give only binary indicators (flagged and not flagged), only a single point for each method is present on the plots. We show results separately for thresholds (left panel) and loadings (right panel). For each continuous measure of fit (MI, EPC, and predicted probability from DNN), we also present the optimal cutpoints (as proposed by Liu 2009).

ROC curves were computed on pulled data from all replications for each simulation condition, investigating the average detection rates among various generated data. For each of the ROC curves, the y-axis corresponds to the true-positive rate (TPR, also called sensitivity), that is, the ratio of correctly classified positive samples (true positives (TP)), to sum of true positives and false negatives (FN, also called type II error). On the x-axis, the false-positive rate (FPR, also called false alarm ratio) might be observed, that is the ratio of false classifications of the positive class (false positives (FP), also known as type I error) and a sum of false positives and true negatives (TN). Those two parameters are useful when describing the performance of classification and detection algorithms, presenting how many positive classes might be detected by making as few mistakes as possible. If one would use the model which makes at most one false alarm (false positive) in 100 predictions, a value of TPR (y-axis) must be read where FPR = 0.01 (x-axis). The TPR here corresponds to the probability of detecting item non-invariance given that it is present, whereas FPR is the probability of detecting it while not appearing, i.e., making a false alarm.

Furthermore, another helpful evaluation metric that comes in handy in pair with the ROC curve is the AUC (area under the curve) score, also referred to as AUROC in this application. AUROC takes values from 0 to 1, where 0 corresponds to the worst and 1 to the best possible performance of the model. It is calculated by simply computing the area between the ROC curve and the x-axis (FPR). Although the standalone value might not be informative enough, they provide a good approximation of the model's performance together with ROC.

**Condition 1: 4 groups 5 items**

Figure 4 presents the results of simulation study 1. Our approach has been proven to be the most effective out of all the methods in this scenario. DNN outperforms other methods both when detecting DIF in thresholds and factor loadings. For all techniques, the thresholds are much easier to detect than factor loadings, which will follow for all studies. A simple procedure based on a logistic model gives high TPR with low FPR for thresholds and yields similar results to DNN for factor loadings. The method based on alignment provides very low FPR with small TPR - this was expected as the method was designed to detect only the largest DIF. MI and EPC yield similar results with a slight advantage for MI.

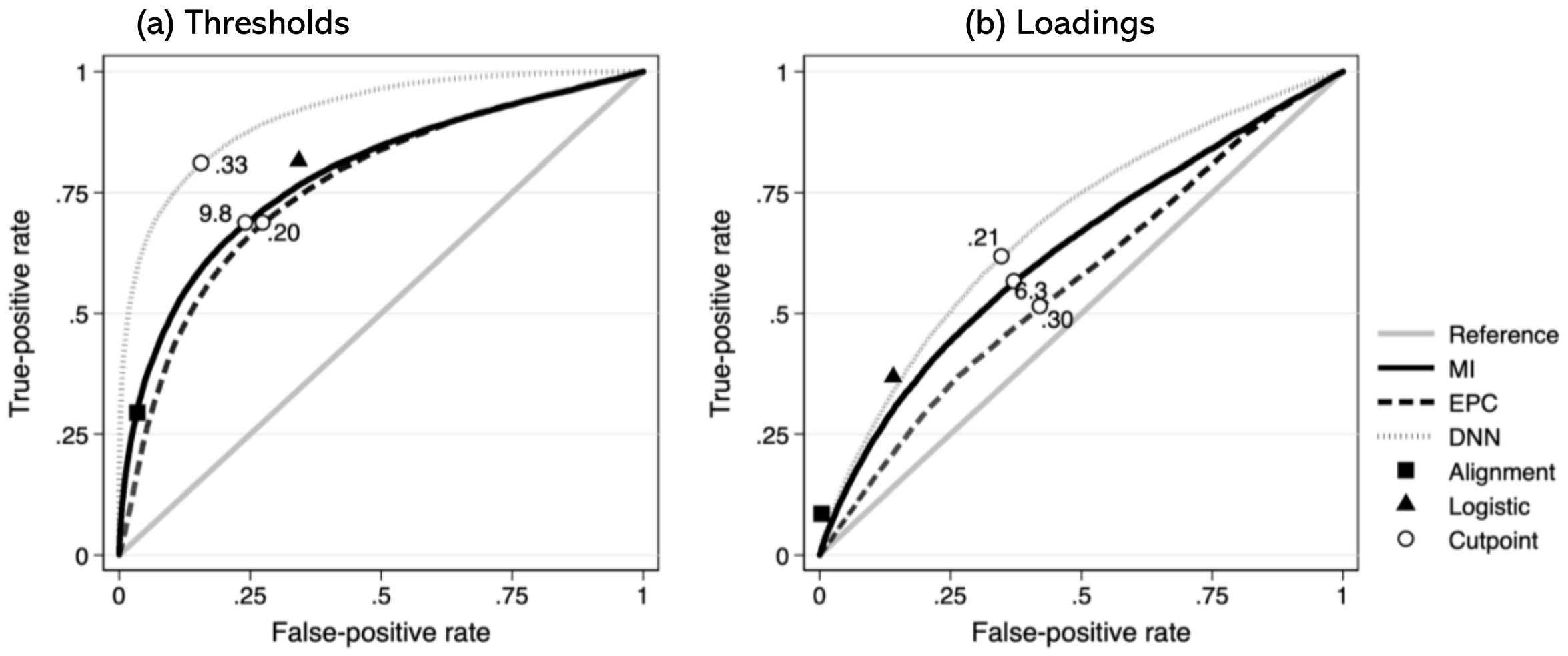

*Figure 4.* ROC curves for simulation study 1 (4 groups and 5 items)

**Condition 2: 10 groups 5 items**

In this simulation study, DNN has the best overall performance as well - it performs much better than MI and EPC, very similar to the method based on logistic regression (on a fixed level of TPR of 0.7). For factor loadings, the performance of DNN is similar to MI, being slightly better for high TPR values but slightly worse when it comes to lower rates. Similar to study 1, the method based on alignment provides very low TPR values.

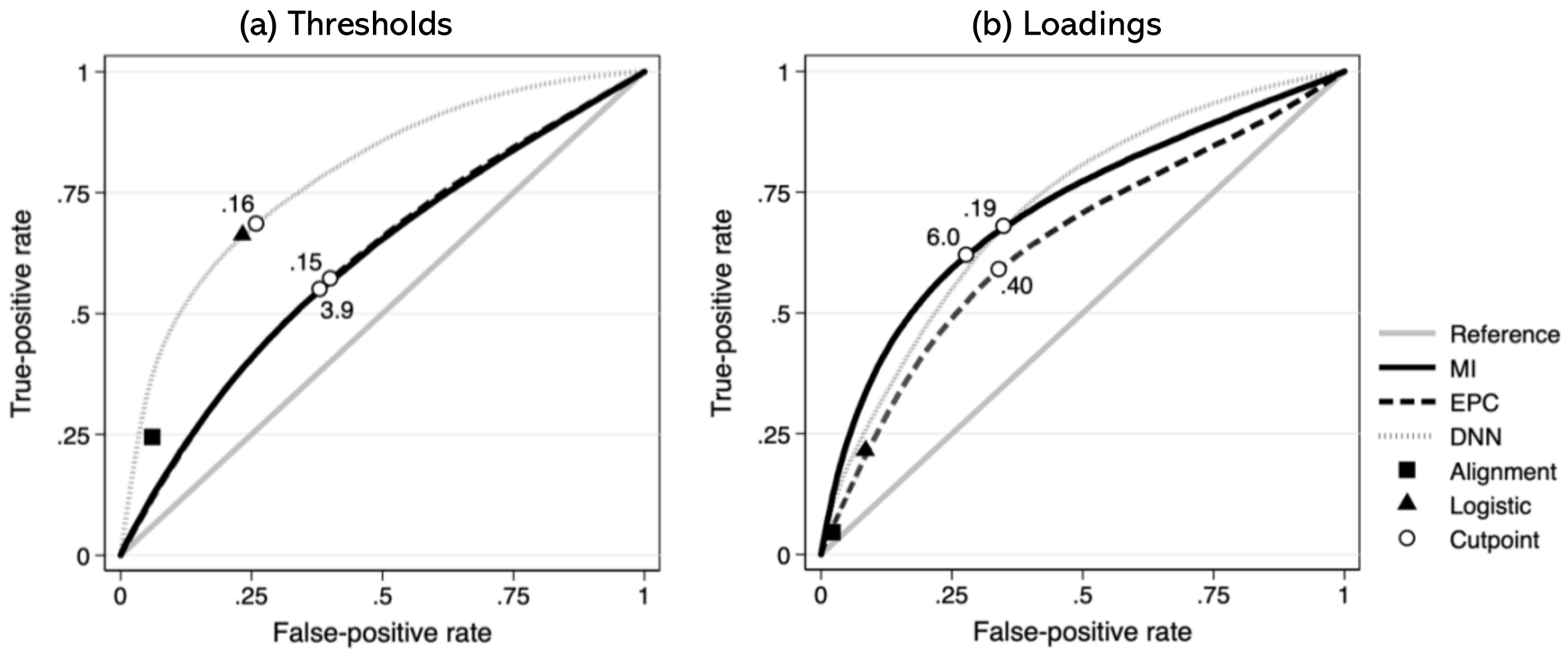

*Figure 5.* ROC curves for simulation study 2 (10 groups and 5 items)

**Condition 3: 4 groups 3 items**

In settings with three items, DNN once again shows its superiority over existing methods, being substantially better than all other approaches both for detection of non-invariant thresholds and non-invariant loadings. The overall detection capability of the latter is relatively poor for all methods, yet still with a notable advantage of DNN.

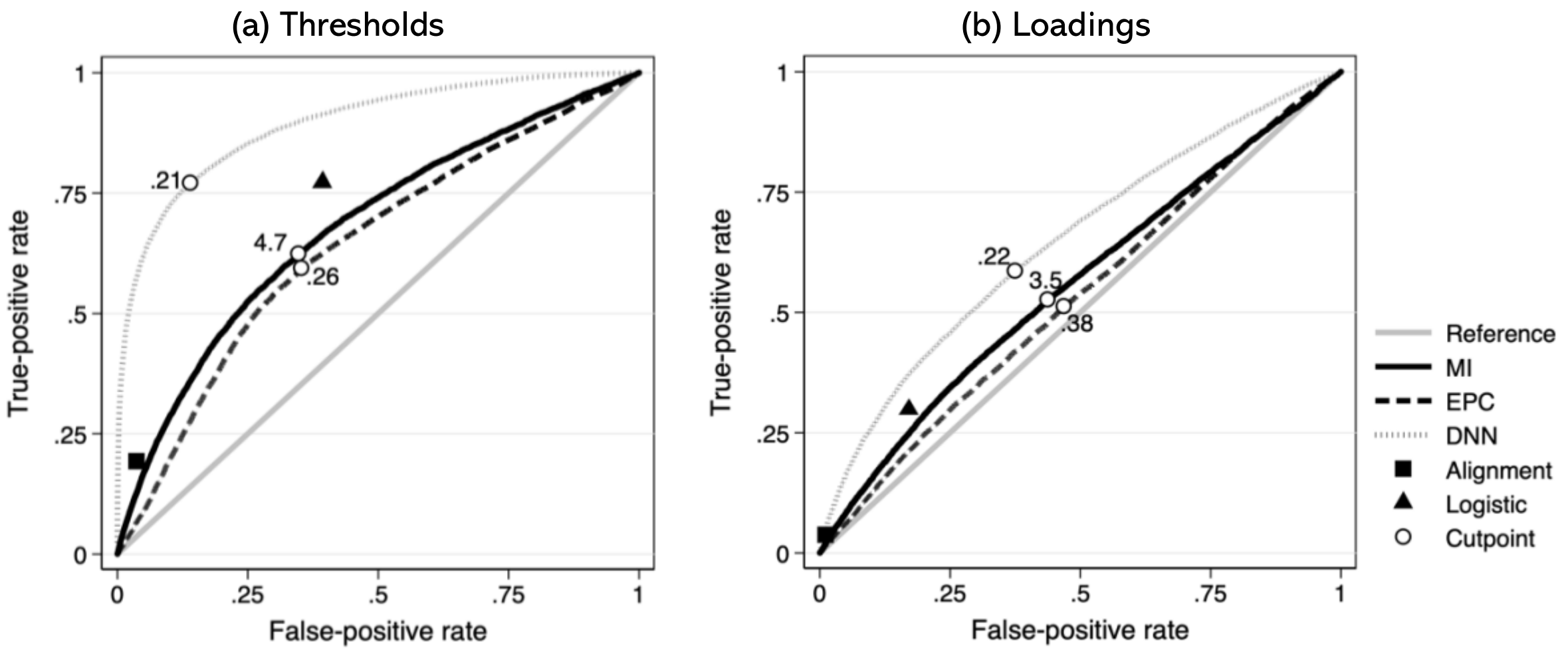

*Figure 6*. ROC curves for simulation study 3 (4 groups and 3 items)

**Condition 4: 10 groups 3 items**

The final simulation results bring no surprise. Similar to previous simulations, DNN performs best with the method based on logistic regression being very close (for one fixed value) for thresholds but substantially worse for factor loadings. Detection based on MI and EPC yields mediocre performance, while detection based on the algorithm used in the alignment approach performs the worst.

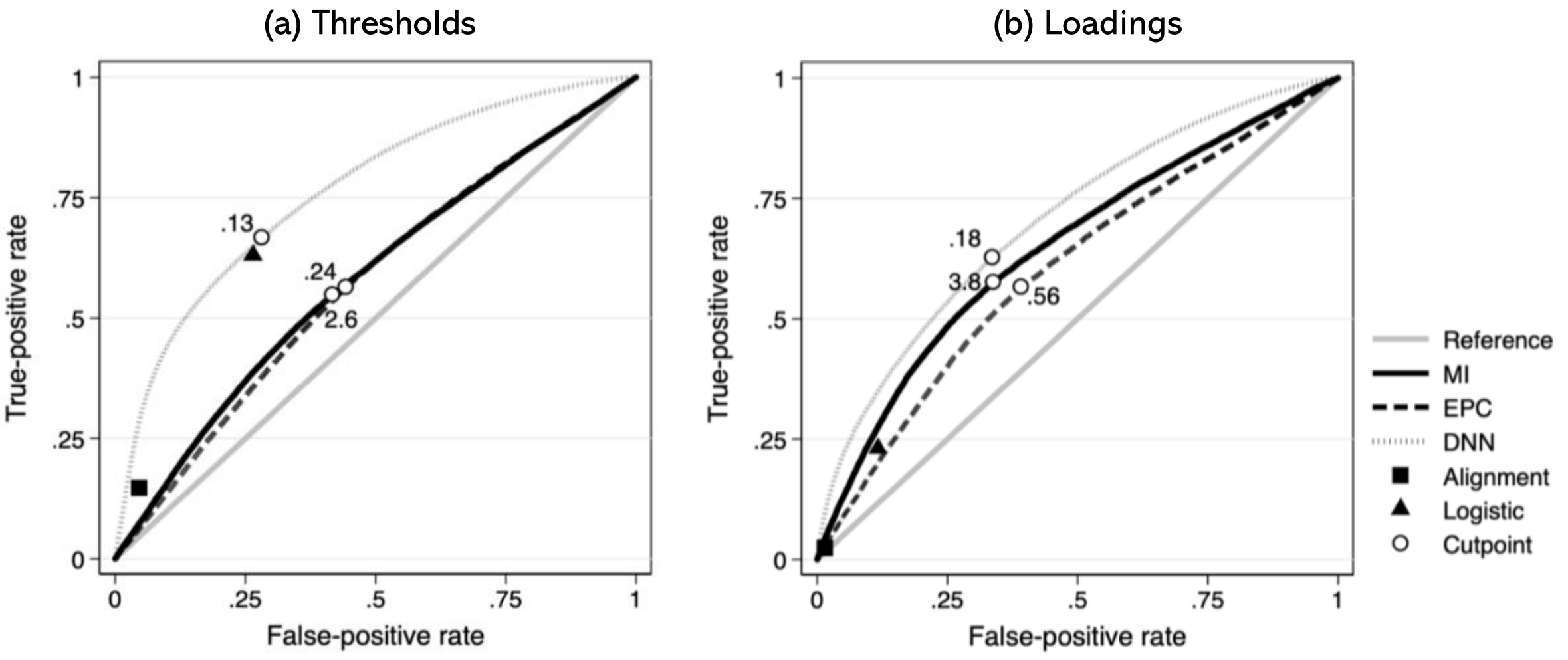

*Figure 7*. ROC curves for simulation study 4 (10 groups and 3 items)

**Summary of the simulation study**

In Table 2, overall comparisons of the performance of the methods are presented by magnitudes of AUROC scores, where the best scores are bolded. Methods that used continuous indicators of DIF could not be directly compared with the methods that provide dichotomous flags (alignment and logistic) because the latter ones are simplified by only one empirical point, and they are not present in the table for this reason.

Table 2 *Results of simulation studies. Area Under ROC (AUROC) score, excluding methods providing dichotomous flag*

| Method | Condition 1 | | Condition 2 | | Condition 3 | | Condition 4 | |
|---|---|---|---|---|---|---|---|---|
| | Thresholds | Loadings | Thresholds | Loadings | Thresholds | Loadings | Thresholds | Loading |
| MI | 0.78 | 0.63 | 0.61 | 0.65 | 0.68 | 0.56 | 0.58 | 0.65 |
| EPC | 0.71 | 0.56 | 0.61 | 0.60 | 0.64 | 0.53 | 0.57 | 0.60 |
| DNN | **0.91** | **0.67** | **0.76** | **0.70** | **0.89** | **0.65** | **0.76** | **0.70** |

*Note.* Best performance bolded.

The AUROC comparisons, remembering its shortcomings, confirms the validity of DNN in simulation settings as it yields the best scores for all studies in terms of this scoring. However, it needs to be noted that alignment and logistic methods were performing similarly to DNN in studies 2 and 4, and there are no AUROC scores for them.

The values of AUC and the graphical representation of ROC curves confirm that the proposed method based on DNN performs better than classical methods and, in most scenarios, substantively outperforms the earlier methods. Our approach provides better detection rates of non-invariance parameters and lowers false detection rates than previously proposed methods for detecting non-invariance parameters in the situation of many groups. The proposed method is the only one that performs highly accurately for both thresholds and factor loadings. Primarily for thresholds, it significantly improves the tested methods for 5- and 3-items scales.

## Empirical applications

We present four examples of applications related to four simulation studies presented earlier, based on ESS data, to show how the proposed methods could be applied to real-life data. The first two empirical examples are based on the 7th round of ESS (ESS 2014). For those, we used five items designed for measuring political participation, where respondents were asked whether they: *Contacted politicians or government officials last 12 months* (i1), *Worked in a political party or action group last 12 months* (i2), *Worked in another organization or association last 12 months* (i3), *Signed petition last 12 months* (i4), and *Boycotted certain products last 12 months* (i5). The items which measure political participation were coded into dichotomous variables, with values 1 "participation" and 0 "no participation". We used the seventh round because a translation error was committed in the Slovenian questionnaire and reported by the ESS consortium – the phrase "*Worked in another organisation or association*" was translated as "*Worked in another political organisation or association*". The item was excluded from the international dataset but was kept in the country-level data. We reintroduced this Slovenian item for our analysis for the purpose of using it as a validation tool. We assume that the good detection method of measurement invariance should flag this item as non-invariant. This is a unique opportunity to test the method in natural experimental settings, which rarely happens in methodological studies on such large-scale data. Additionally, the scale composed of these items was proven to be configural invariance (Koc 2021), showing strong unidimensional structure across all ESS countries.

For examples three and four, we used the most recent ESS data from round 9 (ESS, 2019) utilizing three items measuring political justice: *Political system in the country ensures everyone fair chance to participate in politics* (i1); *Government in the country takes into account the interests of all citizens* (i2); *Decisions in-country politics are transparent* (i3). If respondents agreed with the statement, their response was coded as 1, if disagreed as 0. In this situation, we do not have "naturally" non-invariant items; however, we would like to show the investigated methods' behavior in those settings. The scale was confirmed to be configural invariance across examined countries (McDonald's omega for each group was higher than 0.7).

We performed analysis on four countries and ten countries mimicking the conditions in the simulation studies for each scale. Therefore we use one of the advantages of the proposed method that is the transferability of previously trained networks. To fit our data to the previously trained model, we slightly modified the dataset. For each example, we sampled 400 respondents for each country, including only respondents without missing data for investigated items. As our simulation study was using previously trained network based on actual ESS data, the first three steps of the proposed approach: Initialization (Step 1), Monte Carlo data generation (Step 2), Deep Neural Networks Learning (Step 3), could be omitted as for application on real data we only needed to perform Deep Neural Networks Predicting and Non-Invariance Flagging (Step 4), which simply boils down to using pre-trained networks from simulations and applying them on real data. We feed networks with full vectors of responses from ESS datasets and predict the probability of non-invariance of parameters based on this complete information. To flag the non-invariance items, we used the same optimal (by Liu's 2009 definition) cut points as in the simulation study.

## Results of empirical study

Table 3 presents the results of study one (left panel) and two (right panel). Countries are ordered alphabetically. Items are numbers corresponding to the description provided in previous sections. The most interesting results concern item 3 for Slovenia (SI), which was miss-translated, and we are expecting that suitable DIF methods should flag this item as non-invariant. In fact, in two scenarios (both with 4 and 10 groups) the threshold of this item was flagged both by DNN and method based on pairwise logistic comparisons. Additionally, in scenario 1, EPC flagged it as a threshold and in scenario 2 as a loading. Although this is not the definitive test of the presented method, such a result goes with expectations and increases our confidence in this approach. Moreover, it seems that the two methods with the best performance in the simulations provide results similar to each other, which once again would be something that we would expect from proper techniques.

Alignment methods have not flagged single items in both studies using four and ten groups. This could be expected as simulation studies show a very low detection rate of this approach. MI and EPC, on the other hand, marked too many items, almost all in study 2, which shows its limited usefulness in real data applications. Both DNN and the alignment-based method flagged much fewer items and seemed to have greater levels of agreement than any other pair of methods.

Table 3. *Empirical results from ESS 7 for four and ten groups using political participation items*

| Study 1 | | | | | | | Study 2 | | | | | | |
|---|---|---|---|---|---|---|---|---|---|---|---|---|---|
| **Country (ISCO)** | Item | **MI** | **EPC** | **ALI** | **LOG** | **DNN** | **Country (ISCO)** | Item | **MI** | **EPC** | **ALI** | **LOG** | **DNN** |
| | | | | | | | CZ | 1 | i | i | | | |
| | | | | | | | CZ | 2 | | i | | | |
| | | | | | | | CZ | 3 | | i | | | |
| | | | | | | | CZ | 4 | | i | | | |
| | | | | | | | CZ | 5 | | i | | | |
| | | | | | | | DE | 1 | b | l | | | |
| | | | | | | | DE | 2 | | l | | | |
| | | | | | | | DE | 3 | l | l | | i | |
| | | | | | | | DE | 4 | | l | | | |
| | | | | | | | DE | 5 | | | | i | |
| | | | | | | | ES | 1 | b | b | | l | |
| | | | | | | | ES | 2 | l | b | | | |
| | | | | | | | ES | 3 | b | b | | | |
| | | | | | | | ES | 4 | | b | | | |
| | | | | | | | ES | 5 | | i | | | |
| | | | | | | | FI | 1 | i | i | | | i |
| | | | | | | | FI | 2 | l | b | | | i |
| | | | | | | | FI | 3 | b | b | | i | i |
| | | | | | | | FI | 4 | l | b | | l | i |
| | | | | | | | FI | 5 | l | b | | i | i |
| | | | | | | | FR | 1 | i | i | | | |
| | | | | | | | FR | 2 | i | i | | | |
| | | | | | | | FR | 3 | i | i | | i | |
| | | | | | | | FR | 4 | i | i | | | i |
| | | | | | | | FR | 5 | l | i | | i | i |
| | | | | | | | GB | 1 | b | i | | i | |
| | | | | | | | GB | 2 | i | i | | | i |
| | | | | | | | GB | 3 | b | b | | i | i |
| | | | | | | | GB | 4 | i | b | | i | i |
| | | | | | | | GB | 5 | | i | | | |
| HU | 1 | | | | b | | HU | 1 | l | b | | b | |
| HU | 2 | b | i | | | | HU | 2 | i | b | | | |
| HU | 3 | | | | i | i | HU | 3 | b | b | | | |
| HU | 4 | | | | i | i | HU | 4 | b | b | | i | i |
| HU | 5 | | | | | | HU | 5 | b | b | | | i |
| NO | 1 | l | b | | | | NO | 1 | | b | | | |
| NO | 2 | b | b | | | i | NO | 2 | i | i | | | |
| NO | 3 | l | b | | i | i | NO | 3 | i | i | | i | |
| NO | 4 | | b | | l | | NO | 4 | i | i | | | |
| NO | 5 | | b | | i | i | NO | 5 | i | i | | | |
| PL | 1 | | b | | i | i | PL | 1 | | i | | | |
| PL | 2 | b | b | | | | PL | 2 | | i | | | |
| PL | 3 | | b | | i | | PL | 3 | | | | | |
| PL | 4 | | b | | | | PL | 4 | | l | | | |
| PL | 5 | | i | | i | | PL | 5 | i | i | | i | |
| SI | 1 | | | | | l | SI | 1 | | i | | i | |
| SI | 2 | | | | | i | SI | 2 | | i | | | |
| **SI** | **3*** | | **i** | | **i** | **i** | **SI** | **3*** | | **l** | | **i** | **i** |
| SI | 4 | | | | | | SI | 4 | | l | | | i |
| SI | 5 | | i | | | | SI | 5 | i | i | | | i |

*Note:* i- Threshold flagged as non-invariant; l – loading flagged as non-invariant; b- both Threshold and loading flagged as non-invariant; * naturally non-invariant item

Table 4 presents results using ESS round 9 and 3 items referring to political justice. Here we do not have any "naturally" non-invariant items. The simulation study was predicting a high level of false positives, which is most probably reflected with a large number of flagged items in all methods, excluding an approach based on alignment, which in simulation studies show very low levels of both true and false-positive rates. The overall agreement between the methods is much smaller, which goes along with the simulation results that suggest an overall low detection rate and a significant portion of false positives for the majority of the examined methods.

Table 4. *Empirical results from ESS round 9 for four and ten groups using social justice items*

| Study 3 | | | | | | | Study 4 | | | | | | |
|---|---|---|---|---|---|---|---|---|---|---|---|---|---|
| **Country (ISCO)** | Item | **MI** | **EPC** | **ALI** | **LOG** | **DNN** | **Country (ISCO)** | Item | **MI** | **EPC** | **ALI** | **LOG** | **DNN** |
| | | | | | | | CZ | 1 | i | i | | i | |
| | | | | | | | CZ | 2 | | | | | l |
| | | | | | | | CZ | 3 | | | | i | |
| | | | | | | | DE | 1 | i | | | | i |
| | | | | | | | DE | 2 | | | | | b |
| | | | | | | | DE | 3 | | | | i | l |
| | | | | | | | ES | 1 | | | | | l |
| | | | | | | | ES | 2 | | | | | |
| | | | | | | | ES | 3 | | | | | b |
| | | | | | | | FI | 1 | | l | | | l |
| | | | | | | | FI | 2 | | l | | | |
| | | | | | | | FI | 3 | | l | | | |
| | | | | | | | FR | 1 | l | l | | | |
| | | | | | | | FR | 2 | l | l | | | |
| | | | | | | | FR | 3 | l | | | | |
| | | | | | | | GB | 1 | | l | | | |
| | | | | | | | GB | 2 | l | l | | | |
| | | | | | | | GB | 3 | | | | | l |
| HU | 1 | b | i | | i | l | HU | 1 | b | b | | i | i |
| HU | 2 | i | | l | | l | HU | 2 | i | l | | | |
| HU | 3 | l | | | i | | HU | 3 | b | b | | i | l |
| NO | 1 | b | i | | i | b | NO | 1 | i | b | | i | b |
| NO | 2 | | b | | | b | NO | 2 | i | l | | | i |
| NO | 3 | b | b | | b | | NO | 3 | i | i | | b | |
| PL | 1 | b | | | i | l | PL | 1 | i | b | | i | l |
| PL | 2 | b | l | | | | PL | 2 | i | l | | | |
| PL | 3 | b | i | | b | l | PL | 3 | i | | | i | |
| SI | 1 | b | i | | i | | SI | 1 | b | b | | i | |
| SI | 2 | b | | | i | b | SI | 2 | b | l | | i | i |
| SI | 3 | b | l | | i | | SI | 3 | b | l | i | i | l |

*Note i*- Threshold flagged as non-invariant; l – loading flagged as non-invariant; b- both Threshold and loading flagged as non-invariant

## Discussion and Conclusions

The novel approach of non-invariance detection based on DNN proposed in this paper has yielded very promising results. Simulation studies show that it performs substantially better in most situations, providing greater balance between false positive and true positive rates, while in all other cases, it performs no worse than classical approaches. We showed that it could be successfully applied to real-life data and that it correctly flags items that were previously known to be ("naturally") non-invariant.

The performed simulations are not without limitations - as in all simulation studies, they are bound to a fixed set of conditions. We have tried to avoid it by carefully picking sets that reflect the real data and most plausible situations. Of course, we are open to forthcoming studies that would like to test our approach in different conditions.

Simulations are a critical part of our approach. In the presented study, DNNs were trained on simulated data and then compared with other approaches using the same conditions for training DNNs. Such an approach is a little tautological - although agreeing with such reasoning, we do not see that this is a necessary problem. In the classical approach, model assumptions are stated directly by model formulation. For our approach, the assumptions need to be delivered by learning data generated using Monte Carlo simulations using some assumptions on the nature of invariance. Therefore, in this paper, we have proven that our approach works best if the set of conditions defined by simulation settings is fulfilled. The different question is whether this approach performs well if our assumptions about the measurement invariance will be violated in real data and how robust the presented approach is for such violations. Similar questions could also be asked for classical approaches. For example, what would happen if we would generate data from other models than CFA/IRT. Those are legitimate questions but, in our opinion, out of the scope that could be provided in one article. This is undoubtedly the direction for further studies.

From the other side, the approach could be criticized that for a successful application, we need to precisely design data for simulation, and the created model will be able to detect only specific types of misfit. As it could be treated as problematic, on the other hand, this is a significant advantage that allows us to specify in a very detailed way what kind of DIF or misspecification should be detected and what rates of error are acceptable for the users.

In our application, we showed that the approach based on DNNs could outperform traditional methods. However, the results still could be considered unsatisfactory by some (especially if we are looking at detection rates for factor loadings). In our application, we used relatively shallow DNN architecture that a desktop computer could handle. Given the ongoing development of computational resources, we might expect improvements in utilizing DNNs for the study of this paper and other problems regarding social sciences.

The presented approach was applied to the detection of DIF (or non-invariance) in the context of many groups and short scales and with the plausible assumption that groups substantially differ in latent means amongst each other. Classical approaches did not optimally handle this task, and our proposition is a substantial improvement in the efficiency of detecting non-invariance in the described context. The presented approach could also be applied to other situations where model miss-specification or local misspecification of the statistical model needs to be detected. This could apply to a wide range of more complicated statistical models where detecting model misspecification or data diagnostic is not trivial e.g. dimensionality analysis, assessing, detecting model misspecification in complex non-linear models (e.g. inflated beta regression), detecting response styles or random-effects model misspecification in a rich class of mixed-effects models.

Models and examples described in this paper could be handled easily by a single researcher on a desktop computer. Moving to more realistic applications, e.g., with 30 or 50 groups, would still be manageable; however, computational time will be counted in days rather than hours. As far as single-use is concerned, such restrictions do not seem too problematic. However, problems may arise, for example, in simulation studies where the waiting time for results is already counted in months. Moreover, the proposed method still needs some high-level knowledge on statistical models (for the simulation part) and machine learning algorithms (for the learning part). Developing the approach based on DNN could involve generating many different examples and providing end-users with pre-trained networks with accessible user interfaces. It might be hard for a single researcher to achieve since more complex networks with a higher number of datasets to learn would require using supercomputers and more specialized knowledge. However, in the context of institutions releasing data from many countries and assuring data quality and comparability, such procedures are already within reach. Once trained, DNN could be reused in different situations, as was shown in our empirical example. Applied researchers could further use DNNs prepared by experts as nowadays, statistical packages or open-source software are widely used. There is no reason why tools based on DNN could not be implemented in standard statistical software. The fundamental change is how the statistical device would be developed. In the traditional way, it is written using specified formal assumptions in proposed approaches learned by loads of artificial and (possibly) true data. The recent development of artificial intelligence based on DNN shows that the second path in some situations could also be effective - DNNs achieve the most spectacular successes in problems that are too complex for regular ANNs to solve, such as the field of image recognition. The current DNN solutions achieve an accuracy of 98 percent in such tasks (Mitchell , 2019). In recent years, deep learning has also made contributions not only to applied research; Baldi and colleagues (2014) demonstrate that DNN could be successfully used for analyzing subatomic particle collisions at high-energy particle colliders, improving the classification metrics by as much as 8% over the best approaches that are currently in use. DNNs are used in large physics laboratories, including the Large Hadron Collider at CERN for classification purposes and anomaly detections in particle data (Bhimji et al., 2018). Several works demonstrate that DNNs provide superior performance in predicting the behavior of molecules in pharmaceutical research (Dahl et al., 2014; Wallach et al., 2015). Another area where DNNs were successfully used is applied mathematics: a good example is the three-body problem, which concerns calculating the exact positions of bodies in the future. Although it is possible to solve using conventional techniques, it requires enormous computational resources. Recently, it was shown that DNNs could provide accurate solutions at a fixed computational cost and up to 100 million times faster than state-of-the-art conventional solvers (Breen et al., 2019). Similar approaches are used in astronomy to model galaxy prosperities (Ravanbakhsh et al., 2017) and galaxy classifications (Kennamer et al., 2018). Also, for investigating the sequence specificities of DNA- and RNA-binding proteins (Alipanahi et al., 2015), or performing DNA sequence classification (Bosco & Di Gangi, 2016), DNN has shown its usefulness. Although an exhaustive review of DNN usage in basic research is beyond the scope of this paper, successful application of DNN in climate modeling (Rasp et al., 2018), material science (Mozaffar et al., 2019), ecology (Browning et al., 2018) and in neuroscience (Marblestone et al., 2016) should be mentioned. Why should it not work for developing better sociological methods and research?

# Deep Neural Networks for Detecting Statistical Model Misspecifications. The Case of Measurement Invariance

(Online Support Material)

## Syntax for data generation in R

```
#<=================================================================
# START FUNCTION FOR GENERATING MATRIX DESING FOR SIMULATIONS
#<=================================================================


RandomDesign<- function(Ngroups, Nitems, maxbias_i, maxbias_f, minbias_i,
minbias_f, mifactora=0, mifactorb=0, bmean=0, bsd=1, sdmin, sdmax) {

  #Ngroups        - Number of groups
  #Nitems         - Number of items
  #maxbias_i      - Maximum bias for intercept/threshold
  #maxbias_f      - Maximum bias for loading/slope
  #minbias_i      - Minimum bias for intercept/thrthreshold
  #minbias_f      - Minimum bias for loading/slope
  #mifactora=0    - For approximate MI bias slope. Default No
  #mifactorb=0    - For approximate MI bias threshold. Default No
  #bmean=0        - Mean for sampling group means
  #bsd=1          - SD for sampling group means
  #sdmin=0        - Minimum for uniform dist. for sampling group SD
  #sdmax=1        - Maximium for uniform dist. for sampling group SD

  #<----------------------------------------------------------Build design

  NgrAffected<-sample(c(1:Ngroups), 1)  #  How many groups
  itemID<-rep(seq(1,Nitems,by=1), Ngroups)
  groupID<-rep(seq(1,Ngroups,by=1), rep(Nitems, Ngroups))
  groupA<-c(rep(rep(0,(Nitems*(Ngroups-NgrAffected)))),
          rep(rep(1,(Nitems*(NgrAffected)))))
  itemA<-c(rep(0,Nitems*Ngroups))
```

```
  Tint <-c(rep(0,Nitems*Ngroups))
  bias_b <-runif(Nitems*Ngroups, minbias_f,  maxbias_i)
  bias_a <-runif(Nitems*Ngroups, minbias_f,  maxbias_f)
  s<- sample(c(-1,1), replace=TRUE, size=Nitems*Ngroups)
  bias_b<-bias_b*s
  bias_a<-bias_a*s
  Tload<-c(rep(0,Nitems*Ngroups))
  x<-sample(c(1:3), Ngroups, replace = TRUE, prob = NULL)
  designM<-as.data.frame(cbind(itemID, groupID, groupA, itemA, x, Tint,
  Tload, bias_b, bias_a))

  #<------------------------------------------------Sample affected items

  NitemsN<-Nitems-1
  startGr=Ngroups-NgrAffected
  for (igr in seq(startGr,Ngroups,by=1)) {
    affected<-sample(seq(1,Nitems,by=1), sample(c(1:NitemsN), 1)  , replace
     = FALSE, prob = NULL)
    for (i in affected) {
      designM$itemA[designM$groupA>=1 & designM$itemID==i &
      designM$groupID==igr] <- 1
    }
  }

  #<------------------------------------------------Sample type of bias

  # if 1 only intercepr
  designM$Tint[designM$x==1] <- 1
  # if 2 only load
  designM$Tload[designM$x==2] <- 1
  # if 3 both int and load
  designM$Tint[designM$x==3] <- 1
  designM$Tload[designM$x==3] <- 1
  # rest unaffected
  designM$Tint[designM$itemA==0] <- 0
  designM$Tload[designM$itemA==0] <- 0
  designM$x<-NULL


  #<--------------------------------------------- GEN true item parameters

  designM$aTrue <- rep(rsnorm(Nitems, mean = 2.946, sd = 1.223, xi = 10)
,Ngroups)
  designM$bTrue <- rep(rsnorm(Nitems, mean = -0.2095, sd = 0.6391, xi =
0.7176) ,Ngroups)
  designM$bTrue[designM$bTrue>=  1] <- 1
  designM$bTrue[designM$bTrue<= -2] <- -2

  designM$aTrue[designM$aTrue>=  7] <- 7
```

```
  designM$aTrue[designM$aTrue<= 1]  <- 1

  designM$a <- designM$aTrue
  designM$b <- designM$bTrue


  #<------------------------------------- Gen approx. MI (if necessary)

    if (mifactora>0) {
    # add aproximate MI
    # prior var(dif)=var(mi1)+ var(mi2)
    designM$a<- designM$a+rnorm(1,0,sqrt(mifactora/2))
    designM$b<- designM$b+rnorm(1,0,sqrt(mifactorb/2))
  }

  #<---------------------------------------------------------- Gen Bias

  designM$a[designM$Tload==1]<-  designM$aTrue[designM$Tload==1]
  designM$b[designM$Tint==1] <-  designM$bTrue[designM$Tint==1]

  designM$a[designM$Tload==1]<-
   designM$a[designM$Tload==1]+designM$bias_a[designM$Tload==1]
  # chage sign when degative
  designM$a[designM$a<0]<-
  designM$aTrue[designM$Tload==1]+abs(designM$bias_a[designM$Tload==1])
  designM$b[designM$Tint==1] <-
  designM$b[designM$Tint==1]+designM$bias_b[designM$Tint==1]

  designM$bias_a[designM$Tload==0]<-  0
  designM$bias_b[designM$Tint==0] <-  0


  #<---------------------------------------------------- Gen laten means

  designM$mean <- rep(rnorm(Ngroups, bmean, bsd ) ,each=Nitems)
  designM$sd <- rep(runif(Ngroups, sdmin , sdmax) ,each=Nitems)
  designM$c1<-NULL
  return(designM)
}



#<==================================================================
# GENERATE DATA
# This is the example for 4 groups with sample size 400 with 5 items
# For data generation we use simIrt function for catIrt library
#<==================================================================
```

```
set.seed(13766)


#<--------------------------------------------------- Set working directory

mainpath<-"..."
setwd(mainpath)

#<------------------------------------  Run for 300 000 reps sample size 400

samp<-400
for (repli in seq(1,  300000   ,by=1)) {
  print(repli)

  # Gen DM
DM<-RandomDesign(Ngroups=4, Ngroups=5, maxbias_i=1, maxbias_f=2, mifactora=0,
mifactorb=0, bmean=0, bsd=0.4, sdmin=0.75 , sdmax=1.25)
# Gen Data
samp<-c(400, 400, 400, 400)

sim.data<-NULL
nr<-0
for (repg in seq(1,  4*5 ,by=5)) {
  nr<-nr+1
  print(nr)
  print(repg)
  starti<-repg
  stopi<-repg+4
  sim.params <- cbind(a = DM$a[starti:stopi], b = DM$b[starti:stopi] , c = 0)
  sim.theta<-rnorm(samp[nr], DM$mean[starti], DM$sd[starti] )
  sim.data1<-simIrt(theta=sim.theta, params=sim.params, mod="brm")
  sim.data<-rbind(sim.data, sim.data1$resp)
}

sim.data.v<-as.vector(sim.data)
sim.data<-as.data.frame(cbind(sim.data, c(rep(1,400), rep(2,400), rep(3,400),
rep(4,400))))


#<===================================================================
# SAVE DATA
#<===================================================================

  out<-cbind(DM$Tint, DM$Tload)
  out.v<-as.vector(out)
```

```
pathdata<-paste0("/Users/artur/study1data4NN/rep/rep",repli, ".csv")
write.csv(sim.data.v,pathdata, row.names = FALSE)

pathdata<-paste0("/Users/artur/study1data4NN/out/out",repli, ".csv")
write.csv(out.v,pathdata, row.names = FALSE)

}
```

## Syntax detecting non-invariance (example using 4 groups ad 3 items)

### MI and EPC from the scalar model (Mplus syntax)

```
Data:
    File is datafilr.dat ;
  Variable:
    Names are i1 i2 i3 group
    Missing are all (-9999) ;
    categorical=i1 i2 i3;
    classes = c(4);
    knownclass = c(group = 1-18);

    analysis:
        TYPE=MIXTURE;
        estimator = MLR;
        model = scalar;
        INTEGRATION=30;
        PROCESSORS=4;

    model:
    %Overall%
        F by i1*1 i2*1 i3*1 ;
        F*1; [F*0];
        %C#1%
        F@1; [F@0];
    output: tech1 modindices;
```

### MI and EPC from the scalar model (Mplus syntax)

```
Data:
    File is datafilr.dat ;
  Variable:
    Names are i1 i2 i3 group
```

```
Missing are all (-9999) ;
categorical=i1 i2 i3;
classes = c(4);
knownclass = c(group = 1-4;

analysis:
    TYPE=MIXTURE;
    estimator = MLR;
    ALIGNMENT=FIXED(1);
    INTEGRATION=30;
    PROCESSORS=4;

model:
%Overall%
    F by i1*1 i2*1 i3*1 ;
output: align tech8;
```

## Logistic Regression (R syntax)

```
sim.data is a data vector with 4 colimnns:

V1-V3 0 items 1,2,3

V4    - grouping variable

alpha <- 0.01


# UNIFRON DIF

#<===============================================================
# Compare all groups
#<===============================================================


sim.data.1.2<-subset(sim.data, V4==1 | V4==2)
u<-difLogistic(sim.data.1.2[,1:3], group = sim.data.1.2[,4], focal.name = 1,
type = "udif", alpha = alpha)
u.1.2<-u$p.value

sim.data.1.3<-subset(sim.data, V4==1 | V4==3)
u<-difLogistic(sim.data.1.3[,1:3], group = sim.data.1.3[,4], focal.name = 1,
type = "udif", alpha = alpha)
u.1.3<-u$p.value

sim.data.1.4<-subset(sim.data, V4==1 | V4==4)
```

```
u<-difLogistic(sim.data.1.4[,1:3], group = sim.data.1.4[,4], focal.name = 1,
type = "udif", alpha = alpha)
u.1.4<-u$p.value

sim.data.2.3<-subset(sim.data, V4==2 | V4==3)
u<-difLogistic(sim.data.2.3[,1:3], group = sim.data.2.3[,4], focal.name = 2,
type = "udif", alpha = alpha)
u.2.3<-u$p.value

sim.data.2.4<-subset(sim.data, V4==2 | V4==3)
u<-difLogistic(sim.data.2.4[,1:3], group = sim.data.2.4[,4], focal.name = 2,
type = "udif", alpha = alpha)
u.2.4<-u$p.value

sim.data.3.4<-subset(sim.data, V4==3 | V4==4)
u<-difLogistic(sim.data.3.4[,1:3], group = sim.data.3.4[,4], focal.name = 3,
type = "udif", alpha = alpha)
u.3.4<-u$p.value


#<==================================================================
# # Majority rule
#<==================================================================

  g1<-NULL
for (ite in seq(1,  3   ,by=1))  {
  i<-p.adjust(c(u.1.2[ite], u.1.3[ite], u.1.4[ite]), method ="BH", n = 3)
  print(i)
  i[i<alpha] <- 1
  i[i<1] <- 0
  i<-sum(i)
  g1<-c(g1,i)
}

g2<-NULL
for (ite in seq(1,  3   ,by=1))  {
  i<-p.adjust(c(u.2.3[ite], u.2.4[ite], u.1.2[ite]), method ="BH", n = 3)
  print(i)
  i[i<alpha] <- 1
  i[i<1] <- 0
  i<-sum(i)
  g2<-c(g2,i)
}


g3<-NULL
for (ite in seq(1,  3   ,by=1))  {
  i<-p.adjust(c(u.1.2[ite], u.2.3[ite], u.3.4[ite]), method ="BH", n = 3)
```

```
  print(i)
  i[i<alpha] <- 1
  i[i<1] <- 0
  i<-sum(i)
  g3<-c(g3,i)
}

g4<-NULL
for (ite in seq(1,  3   ,by=1))  {
  i<-p.adjust(c(u.2.4[ite], u.3.4[ite], u.1.4[ite]), method ="BH", n = 3)
  print(i)
  i[i<alpha] <- 1
  i[i<1] <- 0
  i<-sum(i)
  g4<-c(g4,i)
}


INT<-c(g1,g2,g3,g4)
logINT<-rep(0,12)
logINT[INT>1] <- 1
# NON-UNIFRON

#<===================================================================
# Compare all groups
#<===================================================================

sim.data.1.2<-subset(sim.data, V4==1 | V4==2)
u<-difLogistic(sim.data.1.2[,1:3], group = sim.data.1.2[,4], focal.name = 1,
type = "nudif", alpha = alpha)
u.1.2<-u$p.value

sim.data.1.3<-subset(sim.data, V4==1 | V4==3)
u<-difLogistic(sim.data.1.3[,1:3], group = sim.data.1.3[,4], focal.name = 1,
type = "nudif", alpha = alpha)
u.1.3<-u$p.value

sim.data.1.4<-subset(sim.data, V4==1 | V4==4)
u<-difLogistic(sim.data.1.4[,1:3], group = sim.data.1.4[,4], focal.name = 1,
type = "nudif", alpha = alpha)
u.1.4<-u$p.value

sim.data.2.3<-subset(sim.data, V4==2 | V4==3)
u<-difLogistic(sim.data.2.3[,1:3], group = sim.data.2.3[,4], focal.name = 2,
type = "nudif", alpha = alpha)
u.2.3<-u$p.value

sim.data.2.4<-subset(sim.data, V4==2 | V4==3)
```

```
u<-difLogistic(sim.data.2.4[,1:3], group = sim.data.2.4[,4], focal.name = 2,
type = "nudif", alpha = alpha)
u.2.4<-u$p.value

sim.data.3.4<-subset(sim.data, V4==3 | V4==4)
u<-difLogistic(sim.data.3.4[,1:3], group = sim.data.3.4[,4], focal.name = 3,
type = "nudif", alpha = alpha)
u.3.4<-u$p.value

#<==================================================================
# # Majority rule
#<==================================================================

  g1<-NULL
for (ite in seq(1,  3   ,by=1))  {
  i<-p.adjust(c(u.1.2[ite], u.1.3[ite], u.1.4[ite]), method ="BH", n = 3)
  print(i)
  i[i<alpha] <- 1
  i[i<1] <- 0
  i<-sum(i)
  g1<-c(g1,i)
}

g2<-NULL
for (ite in seq(1,  3   ,by=1))  {
  i<-p.adjust(c(u.2.3[ite], u.2.4[ite], u.1.2[ite]), method ="BH", n = 3)
  print(i)
  i[i<alpha] <- 1
  i[i<1] <- 0
  i<-sum(i)
  g2<-c(g2,i)
}


g3<-NULL
for (ite in seq(1,  3   ,by=1))  {
  i<-p.adjust(c(u.1.2[ite], u.2.3[ite], u.3.4[ite]), method ="BH", n = 3)
  print(i)
  i[i<alpha] <- 1
  i[i<1] <- 0
  i<-sum(i)
  g3<-c(g3,i)
}

g4<-NULL
for (ite in seq(1,  3   ,by=1))  {
  i<-p.adjust(c(u.2.4[ite], u.3.4[ite], u.1.4[ite]), method ="BH", n = 3)
  print(i)
```

```
  i[i<alpha] <- 1
  i[i<1] <- 0
  i<-sum(i)
  g4<-c(g4,i)
}


LOA<-c(g1,g2,g3,g4)
logLOA<-rep(0,12)
logLOA[LOA>1] <- 1
```

## DNN (Python/Keras)

```
# this code shows how to declare a DNN model used for this study
# please note that the input and output shapes vary depending on the
# experiment, however the architecture and the optimizer and loss
# function are the same across all of them

# import necessary packages
from tensorflow import keras
from tensorflow.keras import optimizers

# vector_len being the length of the input, i.e. the labels
# output_vector_len being the length of the output, i.e. the features

# create the DNN model
model = keras.Sequential([

    keras.layers.Dense(8000, activation='relu', input_shape=(vector_len,)),
    keras.layers.BatchNormalization(),

    keras.layers.Dense(8000, activation='relu'),
    keras.layers.Dropout(0.5),

    keras.layers.Dense(8000, activation='relu'),
    keras.layers.Dropout(0.5),

    keras.layers.Dense(output_vector_len, activation='sigmoid')
])

# define the optimizer for training
sgd = optimizers.SGD(learning_rate=0.1, momentum=0.8, nesterov=True)

# compile the model, use binary_crossentropy as loss function
model.compile(optimizer=sgd,
```

```
              loss='binary_crossentropy',
              metrics=['accuracy'])

# given X_train is the portion of data for training and X_test is the
# portion for testing (attributes), and also y_train and y_test are
# the features to train the model on and test it respectively, we
# may call the training function as follows:

model.fit(X_train, y_train)

# after the model has been trained, we may produce the prediction for the
# test set. Remember, that this will produce the vector of probabilities
# since we used the sigmoid activation function

y_pred = model.predict(X_test)

# and lastly having function compute_metrics(y_true, y_pred) which
# would compute an arbitrary metric of prediction skill, e.g.
# accuracy, sensitivity etc. Remember that for many metrics we
# first need to binarize the output. We may then call it as:

compute_metrics(y_test, y_pred)
```